%%%%%%%%%%%%%%%%%%%%%%%%%%%%%%%%%%%%%%%%%%%%%%%%%%%%%%%%%%%%%%%%%%%%%%%%%%%%%%%%%%%%%%%%%%%%%%

\input phyzzx

\def\dplus{=\hskip-5pt \raise 0.7pt\hbox{${}_\vert$} ^{\phantom 7}}
\def\dplusup{=\hskip-5.1pt \raise 5.4pt\hbox{${}_\vert$} ^{\phantom 7}}

%%%%%%%%%%%%%%%%%%%%%%%%%%%%%%%%%%%%%%%%%%%%%%%%%%%%%%%%%%%%%%%%%%
\def\dplus{=\hskip-4.8pt \raise 0.7pt\hbox{${}_\vert$} ^{\phantom 7}}

\def\pmb#1{\setbox0=\hbox{#1} \kern-.025em\copy0\kern-\wd0
\kern0.05em\copy0\kern-\wd0 \kern-.025em\raise.0433em\box0}

%%%%%%%%%%%%%%%%%%%%%%%%%%%%%%%%%%%%MACROS%%%%%%%%%%%%%%%%%%%%%%%%%%%%%%%%%%%%%%%%%%

\font\mybb=msbm10 at 10pt

\def\bb#1{\hbox{\mybb#1}}

\def\bT{\bb {T}}
\def\bR{\bb {R}}
\def\bE{\bb {E}}

\sequentialequations

%%%%%%%%%%%%%%%%%%%%%%%%%%%%%%%%%%%%%%%%%%%%%%%%%%%%%%%%%%%%%%%%%%%%
\REF\BST{E. Bergshoeff, E. Sezgin and P.K. Townsend, Phys. Lett. {\bf 189B} (1987) 75;
Ann. Phys. (N.Y.) {\bf 185} (1988) 330.}
\REF\DS{M.J. Duff and K.S. Stelle, Phys. Lett. {\bf 253B} (1991) 113.}
\REF\guv{R. G\"uven, Phys. Lett. {\bf 276B} (1992) 49.}
\REF\townhull{C.M. Hull and P.K. Townsend, Nucl. Phys. {\bf B438} (1995) 109.}
\REF\PKT{P.K. Townsend, Phys. Lett. {\bf 350B}, (1995) 184.}
\REF\JHS{J.H. Schwarz, {\it The Power of M-theory}, hep-th/9510086, {\it M-theory extensions of
T-duality} hep-th/9601077.}
\REF\HW{P. Ho{\v r}ava and E. Witten, Nucl. Phys. {\bf B460} (1996) 506.}
\REF\SBB{K. Becker, M. Becker and A. Strominger, Nucl. Phys. {\bf B456} (1995) 130.}
\REF\Wit{E. Witten, {\it Fivebranes and M-theory on an Orbifold}, hep-th/9512219.}
\REF\BB{K. Becker and M. Becker, {\it Boundaries in M-theory}, hep-th/9602071.}
\REF\ram{J. Rahmfeld, Phys. Lett. {\bf B372} (1996) 198.}
\REF\duff{M.J. Duff and J. Rahmfeld, Phys. Lett. {\bf 345} (1995) 441; M.J.
Duff, J.T. Liu and J. Rahmfeld, Nucl. Phys. {\bf B459} (1996) 125.}
\REF\tseycv{M. Cveti\v c and A.A. Tseytlin, Phys. Lett. {\bf 366B} (1996) 95.}
\REF\sen{A. Sen, Phys. Rev. {\bf D53} (1996) 2874.}
\REF\douglas{M. Douglas, {\it Branes within Branes}, hep-th/9512077.}
\REF\Strom{A. Strominger, {\it Open p-branes}, hep-th/9512059.}
\REF\PKTb{P.K. Townsend, {\it D-branes from M-branes}, hep-th/9512062.}
\REF\Tseytlin{A. A. Tseytlin, {\it Selfduality of Born-Infeld action and Dirichlet 3-brane of type
IIB superstring theory}, hep-th/9602064.}
\REF\greengut{M.B. Green and M. Gutperle, {\it Comments on three-branes}, hep-th/9602077.}
\REF\AT{E.R.C. Abraham and P.K. Townsend, Nucl. Phys. {\bf B351} (1991) 313.}
\REF\memdy{J.M. Izquierdo, N.D. Lambert, G. Papadopoulos and P.K. Townsend, Nucl. Phys. {\bf B460}
(1996) 560.}
\REF\ortin{R. R. Khuri and T. Ort\'\i n, {\it Supersymmetric Black Holes in N=8 Supergravity},
hep-th/9512177.}
\REF\GGT{G.W. Gibbons, G.T. Horowitz and P.K. Townsend, Class. Quantum Grav. {\bf 12} (1995) 297.}
\REF\MP{R.C. Myers and M.J. Perry, Ann. Phys. (N.Y.) {\bf 172} (1986) 304.}
\REF\lupo{ H. L\"u and C.N. Pope, {\it p-brane solutions in maximal
supergravities}, hep-th/9512012.}

%%%%%%%%%%%%%%%%%%%%%%%%%%%%%%%%%%%%TITLE PAGE%%%%%%%%%%%%%%%

\Pubnum{ \vbox{ \hbox{R/96/12} \hbox{hep-th/9603087}} }
\pubtype{}
\date{March, 1996}

\titlepage

\title{Intersecting M-branes}
\author{G. Papadopoulos and P.K. Townsend }
\address{D.A.M.T.P
 \break University of Cambridge\break
         Silver Street \break Cambridge CB3 9EW }
%\andauthor{}
%\address{}

\abstract {We present the magnetic duals of G\"uven's electric-type solutions 
of D=11 supergravity preserving $1/4$ or $1/8$ of the D=11 supersymmetry. We 
interpret the electric  solutions as $n$ orthogonal intersecting membranes and
the magnetic solutions as $n$ orthogonal intersecting 5-branes, with $n=2,3$;
these cases obey the general rule that $p$-branes can self-intersect on
$(p-2)$-branes. On reduction to $D=4$ these solutions become electric or
magnetic dilaton black holes with dilaton coupling constant $a=1$ (for
$n=2$) or $a=1/\sqrt{3}$ (for $n=3$). We also discuss the reduction to D=10.}

\vskip 1cm

\endpage

\pagenumber=2

%%%%%%%%%%%%%%%%%%%%%%%%%%%%%%%INTRODUCTION%%%%%%%%%%%%%%%%%%%%%%%%%%%%%%%%
\chapter{Introduction}

There is now considerable evidence for the existence of a consistent
 supersymmetric quantum theory 
in 11 dimensions (D=11) for which the effective field theory is D=11 
supergravity. This theory, which
goes by the name of M-theory, is possibly a supermembrane theory [\BST]; 
in any case, the membrane
solution of D=11 supergravity [\DS], and its magnetic-dual 5-brane solution [\guv], (which we refer
to jointly as `M-branes') play a central role in what we currently understand about M-theory and its
implications for non-perturbative superstring theory (see, for
example, [\townhull,\PKT,\JHS,\HW,\SBB,\Wit,\BB]). It is therefore clearly of importance to gain a fuller
understanding of {\it all} the $p$-brane-like solutions of D=11 supergravity. 

For example, it was shown by G\"uven [\guv] that the membrane solution of [\DS] is
actually just the first member of a set of three electric-type solutions parametrized, in the 
notation of this paper, by the integer $n=1,2,3$. These solutions are
$$
\eqalign{
ds^2_{(11)}&=-H^{-{2n\over3}}dt^2+H^{n-3\over 3}ds^2(\bE^{2n})+H^{n\over3}ds^2(\bE^{10-2n})
\cr
F_{(11)}&=-3\, dt\wedge dH^{-1}\wedge J\ ,}
\eqn\innine
$$
where $H$ is a harmonic function on $\bE^{10-2n}$ with point singularities, $J$ is a K\"ahler form
on $\bE^{2n}$ and $F_{(11)}$
is the 4-form field strength of D=11 supergravity. The proportion of the D=11 supersymmetry preserved
by these solutions is $2^{-n}$, i.e. ${1\over2},{1\over4}$ and ${1\over8}$, respectively. The $n=1$
case is the membrane solution of [\DS]. We shall refer to the $n=2$ and $n=3$ cases,
which were interpreted in [\guv] as, respectively, a 4-brane and 6-brane, as the `G\"uven
solutions'. Their existence has always been something of a mystery since D=11 supergravity does not
have the five-form or seven-form potentials that one would expect to couple to a 4-brane or a
6-brane. Moreover, unlike the membrane which has a magnetic dual 5-brane, there are no known
magnetic duals of the G\"uven solutions. 

In our opinion, the $p$-brane interpretation given by G\"uven to his electric $n=2,3\;$ solutions
is questionable because of the lack of $(p+1)$-dimensional Poincar\'e invariance
expected of such objects. This is to be contrasted with the $n=1$ case, for which the solution
\innine\ acquires a 3-dimensional Poincar\'e invariance appropriate to its membrane interpretation.
In this paper we shall demystify the G\"uven solutions by re-interpreting them as {\it orthogonally
intersecting membranes}. We also present their magnetic duals which can be interpreted as
orthogonally intersecting 5-branes. The latter are new magnetic-type solutions of D=11 supergravity
preserving, respectively, ${1\over4}$ and ${1\over8}$ of the D=11
supersymmetry.  A novel feature of these solutions is that they involve the
intersection of D=11 fivebranes on 3-branes.  We shall argue that this is an
instance of a general rule: {\it $p$-branes can self-intersect on
$(p-2)$-branes}.

Particle solutions in four dimensions (D=4) can be obtained from M-brane solutions in D=11 by
wrapping them around 2-cycles or 5-cycles of the compactifying space. This is particularly simple in
the case of toroidal reduction to D=4. In this case, wrapped membranes and 5-branes can be
interpreted [\townhull] as, respectively, electric and magnetic $a=\sqrt{3}$ extreme black
holes (in a now standard terminology which we elaborate below). Here we show that G\"uven's 
solutions, and their magnetic duals, have a D=4 interpretation as either $a=1$ (for $n=2$) or 
$a=1/\sqrt{3}$ (for $n=3$) extreme electric or magnetic black holes. 
This D=11 interpretation of the $a=1,1/\sqrt{3}$ extreme black holes in D=4 is 
in striking accord with a recent interpretation [\ram] of them (following earlier
suggestions [\duff], and using results of [\tseycv]) as bound states at threshold of two
(for $a=1$) or three (for $a=1/\sqrt{3}$) $a=\sqrt{3}$ extreme black holes.  

Rather than reduce to D=4 one can instead reduce to D=10 to find various 
solutions of IIA
supergravity representing intersecting $p$-branes. We shall
 briefly mention these at the conclusion of
this paper. There is presumably an overlap here with the discussion of
intersections [\sen] and the `branes within branes'
[\douglas,\Strom,\PKTb,\Tseytlin,\greengut] in the context of D-branes, but we
have not made any direct comparison.  The general problem of intersecting super
p-branes was also discussed in  [\AT] in the context of flat space extended
solitons.  We must also emphasize that the D=11 supergravity solutions we
discuss here have the interpretation we give them only after an integration
over the position of the intersection in the `relative transverse space'; we
argue that this is appropriate for the interpretation as extreme black holes in
D=4.

%%%%%%%%%%%%%%%%%%%%%%%%%%%%%chapter2%%%%%%%%%%%%%%%%%%%%%%%%%%%%%%%%%%%%

\chapter{Intersecting $p$-branes}

We begin by motivating our re-interpretation of the D=11 supergravity 
solutions \innine. The
first point to appreciate is that infinite planar $p$-branes, 
or their parallel multi $p$-brane
generalizations, are not the only type of field configuration for which one can hope to find  
static solutions. Orthogonally intersecting $p$-branes could also be static. The simplest case is that
of pairs of orthogonal $p$-branes intersecting in a $q$-brane, $q<p$. The next simplest case is three
$p$-branes having a {\it common} $q$-brane intersection. Here, however, there is already a 
complication: one must consider whether the intersection of any two of the three $p$-branes is also a
$q$-brane or whether it is an $r$-brane with $r>q$ (we shall encounter both cases below). There are
clearly many other possibilities once one considers more than three intersecting $p$-branes, and even
with only two or three there is the possibility of intersections of orthogonal $p$-branes for
different values of $p$. A limiting case of orthogonal intersections of $p$-branes occurs when
one $p$-brane lies entirely within the other. An example is the D=11 solution of [\memdy]
which can be interpreted as a membrane lying within a 5-brane. For the purposes of this paper,
orthogonal intersections of two or three $p$-branes for the same value of p will suffice.

Consider the case of $n$ intersecting $p$-branes in D-dimensions
 for which the common intersection is a $q$-brane, with worldvolume coordinates
$\xi^\mu$, $\mu=0,1,\dots,q$. The tangent  vectors to the $p$-branes'
worldvolumes  that are {\it not} tangent to the $q$-brane's worldvolume span a
space $V$, which we call the `relative transverse space';  we denote its
coordinates by $x^a$, $a=1,\dots, \ell$, where $\ell={\rm dim} V$. Let $y$
denote the coordinates of the remaining `overall transverse space' of dimension
$D-q-\ell$. The D-dimensional spacetime metric for a system of 
static and   {\it
orthogonal} $p$-branes intersecting in a $q$-brane should take the form
$$
ds^2= A(x,y)d\xi^\mu d\xi^\nu\eta_{\mu\nu} + B_{ab}(x,y) dx^a dx^b
+ C_{ij}(x,y)dy^idy^j\ .
\eqn\anewtwo
$$
Note the $(q+1)$-dimensional Poincar\'e invariance. We also require that
$A\rightarrow 1$, and that $B,C$ tend to the identity matrices, as 
$|y|\rightarrow \infty$, so the metric is asymptotic to the D-dimensional
 Minkowski metric in this limit.

A metric of the form \anewtwo\ will have a standard interpretation as $n$
intersecting $p$-branes only if the coefficients $A,B,C$ functions are such
that the metric approaches that of a single $p$-brane as one goes to infinity
in $V$ while remaining a finite distance from one of the $n$ $p$-branes.  The
G\"uven solutions   \innine\ do not have this property because they are
translation invariant along directions in $V$.  Specifically, they are
special cases of  \anewtwo\ of the form 
$$
ds^2 =A(y)d\xi^\mu d\xi^\nu \eta_{\mu\nu} + B(y)dx^a dx^b\delta_{ab}
 + C(y)dy^i dy^j\delta_{ij}\ .
\eqn\newtwo
$$
Because of the translational invariance in $x$ directions, the energy density
is the same at every point in $V$ for fixed $y$.  However, the translational
invariance allows us to periodically identify the $x$ coordinates, i.e. to take
$V=\bT^\ell$.  In this case, the metric \newtwo\ could be viewed as that of a
$q$-brane formed from the intersection of $p$-branes after averaging over the
intersection points in $V$.  If we insist that the $p$-branes have zero
momentum in $V$-directions orthogonal to their $q$-brane, then this averaging
is an immediate consequence of quantum mechanics.  This delocalization effect
should certainly be taken into account when the size of $V$ is much smaller
than the scale at which we view the dynamics in the $y$ directions, i.e. for
scales at which the effective field theory is $(D-\ell)$-dimensional.  The
$q$-brane solution of this effective field theory can then be lifted to a
solution of the original D-dimensional theory; this solution will be of the
form \newtwo.

Thus, metrics of the form \newtwo\ can be interpreted as those of $p$-branes
intersecting in a common $q$-brane.  However, the solution does not determine,
by itself, the combination of $p$-branes involved.  That is, when interpreted
as a $q$-brane intersection of $n_\alpha$ $p_\alpha$-branes (for
$\alpha=1,2,\dots$) the numbers $(n_\alpha, p_\alpha)$ are not uniquely
determined by the numbers $(D,q,\ell)$.  For example, the $n=2$ G\"uven
spacetime could be interpreted as intersections at a point of (i) 4 strings,
or (ii) 2 strings and one membrane or (iii) a 0-brane and a 4-brane or (iv) 2
membranes.  Additional information is needed to decide between these
possibilities.  In the context of M-theory, most of this additional information
resides in the hypothesis that the `basic' $p$-branes are the M-branes (i.e. the
membrane and 5-brane), where `basic' means that all other $p$-branes-like
objects are to be constructed from them via orthogonal intersections, as
described above.  There is also additional information coming from the form of
the 4-form field strength, which allows us to distinguish between electric,
magnetic and dyonic  solutions.  With this additional information, the
intersecting
$p$-brane interpretation of the $n=2,3$ G\"uven solutions is uniquely that of 2
or 3 intersecting membranes.

It is convenient to consider the G\"uven solutions cases as special
cases of 
$n$ $p$-branes in D
dimensions pairwise intersecting in a common $q$-brane, i.e. $\ell= n(p-q)$. 
To see what to expect of
the magnetic duals of such solutions it is convenient to make a periodic
identification of the
$x$-coordinates in \newtwo, leading to an interpretation of this 
configuration as a $q$-brane in
$d\equiv D-n(p-q)$ dimensions. The magnetic dual of a $q$-brane in 
d dimensions is a $\tilde
q$-brane, where $\tilde q=d-q-4$. We must now find an interpretation
 of this $\tilde q$-brane as an
intersection of $n$ $\tilde p$-branes in D-dimensions, where $\tilde p=D-p-4$. 
The consistency of
this picture requires that the dimension of the space $\tilde V$ spanned
by vectors tangent to the
$\tilde p$-branes' worldvolumes that are {\it not} tangent to the
 $\tilde q$-brane's worldvolume be
$D-d=D-n(p-q)$. This is automatic when $n=2$ (but not when $n>2$). 
As an example, consider the $n=2$
G\"uven solution, interpreted as two orthogonal membranes with a 0-brane
 intersection. Periodic
identification of the $x$-coordinates leads to a particle-like solution
 in an effective D=7
supergravity theory. A particle in D=7 is dual to a 3-brane. This 3-brane
 can now be interpreted as
the intersection of two 5-branes. The vectors tangent to the
5-branes' worldvolumes that are {\it not} tangent to the 3-brane's
 worldvolume span a
four-dimensional space, so the total dimension of the spacetime
 is $7+4=11$, as required. 

Consider now the $n=3$ G\"uven solution, interpreted as three
 orthogonal membranes intersecting at
a common 0-brane. Periodic identification of the $x$-coordinates now
 leads to a particle-like
solution in an effective D=5 supergravity theory. A particle is dual to a
 string in D=5, so we
should look for a solution in D=11 representing three orthogonal 5-branes
 whose common intersection
is a string. The dimension of the space $\tilde V$ spanned by the vectors
 tangent to the 5-branes'
worldvolumes that are {\it not} tangent to the string's worldsheet depends
 on whether the common
intersection of all three 5-branes is also the intersection of any pair. 
If it were then $\tilde V$
would take its maximal dimension, $3(5-1)=14$, leading to a total spacetime
 dimension of $5+14=19$.
Since this is inconsistent with an interpretation in D=11, we conclude that
 the pairwise
intersection of the three 5-branes must be a $q$-brane with $q>1$. In fact,
 the consistent choice is
$q=3$, i.e. each pair of 5-branes has a 3-brane intersection and the three
 3-branes themselves
intersect in a string\foot{A useful analogy is that of three orthogonal
 planes in $\bE^3$ which
intersect pairwise on a line. The three lines intersect at a point.}.
In this case $\tilde V$ has
dimension six, leading to a total spacetime dimension of eleven. 

Note that all the cases of intersecting $p$-branes which we have argued should
occur in M-theory have the property that $p$-brane pairs (for the same value of
$p$) intersect on $(p-2)$-branes.  Specifically, we have argued that 2-branes can
intersect on 0-branes, that 5-branes can intersect on 3-branes and that these
3-brane intersections can themselves intersect on 1-branes.  We shall conclude
this section by explaining why we believe that this is a general rule, i.e.
{\it $p$-branes can self-intersect on $(p-2)$-branes}.

Recall that the possibility of a membrane having a boundary on a 5-brane 
[\Strom,\PKTb] arises from the fact that the 5-brane worldvolume contains a
2-form potential which can couple to the membrane's string boundary.  The same
argument does not obviously apply to {\it intersections} but it is plausible
that it does, at least for those cases in which it is possible to view the
$q$-brane intersection within a given $p$-brane as a dynamical object in its
own right.  Thus, it is reasonable to suppose that a condition for a $p$-brane
to support a $q$-brane intersection is that the $p$-brane worldvolume field
theory includes a $(q+1)$-form potential to which the $q$-brane can couple.  We
now observe that $p$-brane worldvolume actions always contain $(D-p-1)$ scalar
fields.  If one of these scalars is dualized then the worldvolume acquires a
$(p-1)$-form potential, which can couple to a $(p-2)$-brane.  Hence the rule
stated above; the freedom of choice of which scalar to dualize corresponds to
the possibility of an energy flow into the $p$-brane, at the intersection, in any
of the directions orthogonal to its worldvolume.

%%%%%%%%%%%%%%%%%%%chapter3%%%%%%%%%%%%%%%%%%%%%%%%%%

\chapter{Magnetic duals of G\"uven solutions} 

We now have sufficient information to find the magnetic duals of the series of electric solutions
\innine\ of D=11 supergravity. They should be of the form \newtwo\ with $q=7-2n$ and they should
preserve some fraction of the D=11 supersymmetry. Solutions that preserve some 
supersymmetry can most easily be found by seeking bosonic backgrounds admitting Killing spinors. The
Killing spinor equation can be found directly from the supersymmetry transformation law for the
gravitino field $\psi_M$  ($M=0,1,2,\dots,10$),  and is 
$$
\Big[D_M + {1\over144}\big(\Gamma_M{}^{NPQR} - 8\delta_M^N
\Gamma^{PQR}\big) F_{NPQR}\Big]\zeta =0\ ,
\eqn\ainthirteen
$$
where $D_M$ is the standard covariant derivative. Solutions $\zeta$ of this equation (if any)
are the Killing spinors of the bosonic background, i.e. the D=11 metric and 4-form field strength
$F_{MNPQ}$. Backgrounds admitting Killing spinors for which the Bianchi identity for $F_{(11)}$ is
also satisfied are automatically solutions of D=11 supergravity. The proportion of the D=11
supersymmetry preserved by such a solution equals the dimension of the space of Killing spinors
divided by 32. 

By substituting an appropriate ansatz for the metric and 4-form into \ainthirteen\ we have found a
series of magnetic solutions parametrised by the integer $n=1,2,3$. These are
$$
\eqalign{
ds^2_{(11)}&=H^{-{n\over3}}(d\xi\cdot d\xi)+ H^{-{n-3\over
3}}ds^2(\bE^{2n})+ H^{2n\over3} ds^2(\bE^3)
\cr
F_{(11)}&=\pm 3 \,\star dH\wedge J \ ,}
\eqn\inthirteen
$$
where $\star$ is the Hodge star of $\bE^3$, $J$ is the K\"ahler form on $\bE^{2n}$ and $n=1,2,3$.
Our conventions for forms are such that 
$$
\eqalign{
J &={1\over 2} J_{ab}dx^a\wedge dx^b\cr
F_{(11)}&={1\over4} F_{MNPR}\, dx^M\wedge dx^N\wedge dx^P\wedge dx^R \ .}
\eqn\innineab
$$
The function H is harmonic on $\bE^3$ with point singularities. Asymptotic
flatness at `overall transverse infinity' requires that $H\rightarrow 1$ there, so that
$$
H=1+\sum_i {\mu_i\over |x-x_i|}\ ,
\eqn\infive
$$
for some constants $\mu_i$. Note that these solutions have an
$8-2n$ dimensional Poincar\'e invariance, as required.

In the $n=1$ case the metric can be written as
$$
ds^2_{(11)} =H^{-{1\over3}}\, d\xi\cdot d\xi + H^{2\over 3}ds^2(\bE^5)
\eqn\fivebrane
$$
which is formally the same as the 5-brane solution of [\guv]. The difference is that the function
$H$ in our solution is harmonic on an $\bE^3$ subspace of $\bE^5$, i.e. our solution is a special
case of the general 5-brane solution, for which $H$ is harmonic on $\bE^5$. The $n=2,3$ cases are
{\it new} solutions of D=11  supergravity with the properties expected from their interpretation as
intersecting 5-branes. The solutions of the Killing spinor equation for the background given by
\inthirteen\ are 
$$
\eqalign{
\zeta&=H^{-{n\over 12}} \zeta_0
\cr
\bar\Gamma_a\zeta_0&=\mp J_{ab}\gamma^*\bar\Gamma^b\zeta_0\ , }
\eqn\infourteen
$$
where $\{\bar\Gamma_a; a=1,\dots,2n\}$ are the (frame) constant D=11 gamma matrices along the
$\bE^{2n}$ directions, $\gamma^*$ is the product of the three constant gamma matrices along
the $\bE^3$ directions and $\zeta_0$ is a constant D=11 spinor. It follows from \infourteen\ that
the number of supersymmetries preserved by the magnetic intersecting 5-brane solutions is $2^{-n}$,
exactly as in the electric case.

%%%%%%%%%%%%%%%%%%%%%%%%%%%%%%%chapter 4%%%%%%%%%%%%%%%%%%%%%%%%%%%%%

\chapter{D=4 Interpretation}

We now discuss the interpretation of the solutions \innine\ and 
\inthirteen\ in D=4. The D=4 field theory obtained by compactifying D=11 supergravity on $\bT^7$
can be consistently truncated to the massless fields of N=8 supergravity. The latter can be
truncated to
$$
I=\int\! d^4x {\sqrt{-g}} \bigg[R-2(\partial\phi)^2-{1\over2} e^{-2a\phi} F^2\bigg]\ ,
\eqn\inone
$$
where $F$ is an abelian 2-form field strength, provided that the scalar/vector coupling constant
$a$ takes one of the values\foot{We may assume that $a\geq 0$ without loss of
generality.}[\townhull,\ortin]
$$
a=\sqrt 3\ , \ 1\ ,\ {1\over \sqrt 3}\ , \  0\ .
\eqn\intwo
$$
The truncation of N=8 supergravity to \inone\ is not actually a consistent one (in the standard
Kaluza-Klein sense) since consistency requires that $F$ satisfy $F\wedge F=0$. However, this
condition is satisfied for purely electric or purely magnetic field configurations, so
purely electric or purely magnetic solutions of the field equations of \inone\ are automatically
solutions of N=8 supergravity, for the above values of $a$. In particular, the static extreme
electric or magnetic black holes are solutions of N=8 supergravity that
preserve some proportion of the N=8 supersymmetry. This proportion is
$1/2,1/4,1/8,1/8$ for $a=\sqrt{3},1,1/\sqrt{3},0$, respectively.

It is known that the membrane and fivebrane solutions of D=11 supergravity have a D=4
interpretation as $a=\sqrt{3}$ extreme black holes. Here we shall extend this result to the
$n=2,3$ cases by showing that the electric solutions \inone\ of D=11 supergravity, and their magnetic
duals \inthirteen\ have a D=4 interpretation as extreme black holes with scalar/vector coupling
$a=\sqrt{(4/n) -1}$. As we have seen, the D=11 solutions for $n=2,3$, electric or magnetic, have a
natural interpretation as particles in D=7 and D=5, respectively. It is therefore convenient to
consider a two-step reduction to D=4, passing by these intermediate dimensions. The $n=3$ case is
actually simpler, so we shall consider it first. We first note that for $a=1/\sqrt{3}$ the action
\inone\ can be obtained from that of simple supergravity in D=5, for which the bosonic fields are
the metric $ds^2_{(5)}$ and an abelian vector potential $A$ with 2-form field strength $F_{(5)}$, by
the ansatz
$$
\eqalign{
ds^2_{(5)}&=e^{2\phi} ds^2+e^{-4\phi} dx_5^2
\cr
F_{(5)}&=F\ ,}
\eqn\inseven
$$
where $ds^2$, $\phi$ and $F$ are the metric and fields appearing in the D=4 action \inone.
Note that this ansatz involves the truncation of the D=4 axion field $A_5$; it is the
consistency of this truncation that requires $F\wedge F=0$. As mentioned above, this does not
present problems in the purely electric or purely magnetic cases, so these D=4 extreme black hole
solutions can be lifted, for $a=1/\sqrt{3}$, to solutions of D=5 supergravity. The
magnetic black hole lifts to the D=5 extreme black multi string solution [\GGT]
$$
\eqalign{
ds^2_{(5)} &= H^{-1} (-dt^2 + dx_5^2) + H^2 ds^2(\bE^3)\cr
F_{(5)} &={}^\star dH \ ,}
\eqn\bstring
$$
where $\star$ is the Hodge star of $\bE^3$ and $H$ is a harmonic function on $\bE^3$
with some number of point singularities, i.e. as in \infive. We get a magnetic $a=1/\sqrt{3}$
extreme black hole by wrapping this string around the $x_5$ direction. 

The electric $a=1/\sqrt{3}$ extreme multi black hole lifts to the following solution of D=5
supergravity:
$$
\eqalign{
ds^2_{(5)}&=-H^{-2} dt^2+H ds^2({\bE^3\times S^1})
\cr
F_{(5)}&=dt\wedge dH^{-1}\ ,}
\eqn\infour
$$
where $H$ is a harmonic function on $\bE^3$. This solution is the `direct' dimensional reduction of
the extreme electrically-charged black hole solution of D=5 supergravity [\MP]. The latter is formally the same
as \infour\ but $\bE^3\times S^1$ is replaced by $\bE^4$ and $H$ becomes a harmonic function on
$\bE^4$.

To make the connection with D=11 we note that the Kaluza-Klein (KK) ansatz
$$
\eqalign{
ds^2_{(11)}&=ds_{(5)}^2+ds^2(\bE^6)
\cr
F_{(11)}&=F_{(5)}\wedge J\ ,}
\eqn\ineighta
$$
where $J$ is a K\"ahler 2-form on $\bE^6$, provides a consistent truncation of D=11 supergravity to the fields of
D=5 simple supergravity. This allows us to lift solutions of D=5 supergravity directly to D=11. It
is a simple matter to check that the D=5 extreme black hole solution lifts to the $n=3$ G\"uven
solution and that the D=5 extreme black string lifts to the magnetic $n=3$ solution of \inthirteen.

The $a=1$ case works similarly except that the intermediate dimension is D=7. The KK/truncation ansatz taking
us to D=7 is 
$$
\eqalign{
ds_{(11)}^2 &= e^{-{4\over3}\phi}d{\hat s}_{(7)}^2 + e^{{2\over3}\phi} ds^2(\bT^4)\cr
F_{(11)} &= F_{(7)}\wedge J \ .}
\eqn\dimreda
$$
where $d{\hat s}^2_{(7)}$ is the string-frame D=7 metric. Consistency of this truncation
restricts $F_{(7)}$ to satisfy $F_{(7)}\wedge F_{(7)}=0$, but this will be satisfied by our
solutions. The ansatz then taking us to D=4 is
$$
\eqalign{
d{\hat s}^2_{(7)}&= d{\hat s}^2+ ds^2(\bT^3)\cr
F_{(7)} &= F\ ,}
\eqn\dimredb
$$
where $d{\hat s}^2= e^{2\phi} ds^2$ is the string-frame D=4 metric.
Combining the two KK ans\"atze, it is not difficult to check that the electric $a=1$
extreme black hole lifts to the $n=2$ G\"uven solution in D=11 and that the magnetic $a=1$ extreme
black hole lifts to the new $n=2$ magnetic D=11 solution of this paper.

%%%%%%%%%%%%%%%%%%%%%%%%%%%%%%%%%%%%%%%%%%%%%%%%%%%%%%%%%%%%%%%%%%%%%%%%
\chapter{Comments}

We have extended the D=11 interpretation of D=4 extreme black hole solutions of N=8 supergravity
with scalar/vector coupling $a=\sqrt{3}$ to two of the other three possible values, namely $a=1$ and
$a=1/\sqrt{3}$. While the $a=\sqrt{3}$ black holes have a D=11 interpretation as wrapped M-branes,
the $a=1$ and $a=1/\sqrt{3}$ black holes have an interpretation as wrappings of, respectively, two or
three {\it intersecting} M-branes. We have found no such interpretation for the $a=0$ case, i.e. extreme
Reissner-Nordstr\"om black holes; we suspect that their D=11 interpretation must involve the gauge
fields of KK origin (whereas this is optional for the other values of $a$).

The solution of D=11 supergravity representing three intersecting 5-branes is essentially the
same as the extreme black string solution of D=5 supergravity. For both this solution and the D=11
5-brane itself the singularities of $H$ are actually coordinate singularities at event horizons.
Moreover, these solutions were shown in [\GGT] to be geodesically complete, despite the existence
of horizons, so it is of interest to consider the global structure of the solution representing two
intersecting 5-branes. For this solution the asymptotic form of $H$ near one of its singularities is
$H\sim 1/r$, where $r$ is the radial coordinate of $\bE^3$. Defining a new radial coordinate $\rho$
by
$r=\rho^3$, we find that the asymptotic form of the metric near $\rho=0$ is
$$
ds^2_{(11)}\sim \rho^2 d\xi\cdot d\xi + {1\over\rho} ds^2(\bE^4) + 9\, d\rho^2 + \rho^2 d\Omega^2
\eqn\commenta
$$
where $d\Omega^2$ is the metric of the unit 2-sphere. `Spatial' sections of this metric, i.e. those
with $d\xi=0$, are topologically $\bE^4\times S^2\times \bR^+$, where $\rho$ is the
coordinate of $\bR^+$. Such sections are singular at $\rho=0$ although it is notable that the volume
element of $\bE^4\times S^2$ remains finite as $\rho\rightarrow 0$.

We have concentrated in this paper on solutions representing intersecting $p$-branes in D=11,
i.e. M-branes, but the main idea is of course applicable to supergravity theories in lower
dimensions. In fact, the intersecting M-brane solutions in D=11 can be used to deduce solutions
of D=10 IIA supergravity with a similar, or identical, interpretation by means of either direct 
or double dimensional reduction. Direct reduction yields solutions of D=10 IIA supergravity with
exactly the same interpretation as in  D=11, i.e.  two (for $n=2$) or three (for $n=3$) membranes
intersecting at a point, in the electric case, and, in the magnetic case, two 5-branes intersecting
at a 3-brane (for $n=2$) or three 5-branes intersecting at a string (for $n=3$).  On the other
hand, double dimensional reduction of the electric D=11 $n>1$ solutions, i.e. wrapping one membrane
around the $S^1$, gives solutions of D=10 N=2A supergravity theory representing either 
a string and a membrane intersecting at a point (for $n=2$) or a string and two membranes
intersecting at a point (for $n=3$). In the magnetic case, the wrapping can be
done in two different ways. One way, which is equivalent to double-dimensional reduction,
is to wrap along one of the relative transverse directions, in which case
the D=10 solutions represent either a 5-brane and a 4-brane intersecting at a 3-brane (for $n=2$) or
two 5-branes and a 4-brane intersecting at a string (for $n=3$). The other way, which might
reasonably be called `triple dimensional' reduction, is to wrap along one of the directions in the
{\it common} $q$-brane intersection, in which case one gets D=10 solutions representing either
two 4-branes intersecting at a membrane (for $n=2$) or three 4-branes intersecting at a point (for
$n=3$). We expect that some of these IIA D=10 solutions will have a superstring description via
Dirichlet-branes. 

Finally, we point out that the solutions \innine\ and \inthirteen\ can
both be generalized to the
case in which $ds^2(\bE^{2n})$ is replaced by any Ricci-flat K\"ahler 
manifold ${\cal M}^n$ of
complex dimension $n$. Examples of
compact manifolds
${\cal M}^n$ for
 $n=1,2,3$ are ${\cal
M}^1=\bT^2$,
${\cal M}^2=K_3$, ${\cal M}^3$ a Calabi-Yau space. The new solutions of
 D=11 supergravity obtained
in this way generalize the corresponding KK vacuum solution of D=11 
supergravity to one
representing an M-brane, or intersecting M-branes, wrapped around cycles in the
the compactifying space.  In any case, it is clear that the results of this
paper are far from complete.  It seems possible that a recent classification
[\lupo] of $p$-brane solutions of maximal supergravities in dimensions $D<11$
might form a basis of a systematic M-theory interpretation, along the lines
presented here, of all p-brane like solutions of D=11 supergravity.

\vskip 0.5cm

\noindent{\bf Acknowledgments:} 
 We would like to thank M.B. Green, J. Gauntlett and E. Witten for helpful
discussions. G.P. is supported by a University Research Fellowship from  the
Royal Society.

\refout

\bye